\documentstyle[11pt,newpasp,twoside,epsf]{article}
\begin{document}
\title{The Deuterium Abundance in QSO Absorption Systems~:
A Mesoturbulent Approach\footnotemark[1]}
\footnotetext[1]{Based on data obtained at the
W. M. Keck Observatory, which is jointly operated by the
University of California, the California Institute of Technology,
and the National Aeronautics and Space Administration.} 
\author{Sergei A. Levshakov}
\affil{Department of Theoretical Astrophysics,
Ioffe Physico-Technical Institute, 194021 St.~Petersburg,
Russia}
\begin{abstract}
A new method, based on simulated annealing technique 
and aimed at the inverse problem in the analysis of 
intergalactic or interstellar complex spectra of
hydrogen and metal lines, is outlined.
We consider the process of line formation in clumpy 
stochastic media accounting for fluctuating velocity and density fields
self-consistently.
Two examples of the analysis of
`H+D'-like absorptions seen at $z_{\rm a} = 3.514$
and 3.378 towards APM 08279+5255 are presented.
\end{abstract}

\section{Introduction}

The cosmological significance of the deuterium abundance
measurements in metal-deficient QSO absorption systems
has been widely discussed in the literature (see e.g. the 
review by Lemoine et al. 1999).
Practical applications of such measurements were
clearly outlined in Tytler \& Burles (1997)~: 
(i) the primordial D/H value gives the density of baryons $\Omega_{\rm b}$
at the time of big bang nucleosynthesis (BBN), 
a precise value for $\Omega_{\rm b}$  might be used
(ii) to determine the fraction
of baryons which are missing,  
(iii) to specify Galactic chemical
evolution, and  
(iv) to test models of high energy physics. 
A measurement of the D/H ratio, 
together with other three light element abundances
($^3$He/H, $^4$He/H, and $^7$Li/H), provides the complete test
of the standard BBN model. 

Deuterium has been reported up-to-now in a few 
QSO Lyman limit systems (LLS), -- the systems
with the neutral hydrogen column densities of 
$N_{{\rm H}\,{\sc i}} \sim 10^{17}-10^{18}$~cm$^{-2}$
(Kirkman et al. 1999).
The difficulties inherent to measurements of D/H in QSO spectra 
are mainly caused by 
the confusion between the D\,{\sc i} line 
(which is always partially blended by the blue wing of the 
saturated hydrogen line) and the numerous neighboring weak lines
of H\,{\sc i} observed in the Ly$\alpha$ forest at redshifts
$z > 2$ (Burles \& Tytler 1998).

Currently, there are two methods to analyse the
absorption spectra~: (i) a conventional Voigt-profile fitting (VPF)
procedure which usually assumes several subcomponents
with their own physical parameters
to describe a complex absorption profile, and 
(ii) a mesoturbulent approach which describes the line formation process
in a continuous medium with fluctuating 
physical characteristics.
It is hard to favor this or that method if both of them provide 
good fitting.
But the observed increasing
complexity of the line profiles with increasing spectral resolution 
gives some preference
to the model of the fluctuating continuous medium.

Here, we set forward a mesoturbulent approach to measure D/H
and metal abundances, which has many advantages over the
standard VPF procedures. 
A brief description of a new Monte Carlo inversion (MCI) 
method is given in this report.
For more details, the reader is referred to
Levshakov, Agafonova, \& Kegel (2000b).

An example of the MCI analysis of two `H+D'-like profiles 
with accompanying metal lines
observed at $z_{\rm a} = 3.514$ and 3.378 towards
the quasar APM 08279+5255  is described.
The high quality spectral data have been
obtained with the Keck-I telescope and the HIRES spectrograph by
Ellison et al. (1999).

\section{The MCI method and results}

The MCI method is based on simulated annealing technique and aimed at
the evaluation both the physical parameters of the gas cloud
and the corresponding velocity and density distributions 
along the line of sight. 
We consider the line formation process in clumpy stochastic 
media with fluctuating velocity and density fields
({\it mesoturbulence}).
The new approach generalizes our previous Reverse Monte Carlo
(Levshakov, Kegel, \& Takahara 1999)
and Entropy-Regularized Minimization 
(Levshakov, Takahara, \& Agafonova 1999)
methods dealing with incompressible turbulence (i.e. the case of
random bulk motions with
homogeneous gas density $n_{\rm H}$ and kinetic temperature $T$).

The main goal is to solve the {\it inverse problem},
i.e. the problem to deduce physical parameters from 
a QSO absorption system. 
The inversion is always an optimization problem in which an
objective function is minimized. 
To estimate a goodness of the minimization 
we used a $\chi^2$ function augmented by a regularization term 
(a penalty function) to stabilize the MCI solutions.
The MCI is a stochastic optimization procedure
and one does not know in advance
if the global minimum of the objective function is reached in
a single run.
Therefore to check the convergency, 
several runs are executed for a given data set 
with every calculation
starting from a random point in the simulation 
parameter box and
from completely random
configurations of the velocity and density fields.
After these runs, the  
median estimation of the model parameters is performed.

Our model supposes a continuous absorbing gas slab of a thickness $L$.
The velocity component along a given line of sight is described 
by a random function in which the velocities in 
neighboring volume elements are correlated with each other. 
The gas is optically thin in the Lyman continuum.
We are considering a compressible gas, i.e. 
$n_{\rm H}$ is also a random function
of the space coordinate, $x$.
Following Donahue \& Shull (1991) and assuming that the ionizing radiation
field is constant,
the ionization of different elements can be described by
one parameter only -- the ionization parameter
$U \propto 1/n_{\rm H}$. Furthermore,
for gas in thermal
equilibrium, Donahue \& Shull give an explicit relation between $U$
and $T$. 
The background ionizing spectrum is taken from
Mathews \& Ferland (1987).

In our computations, the continuous random functions $v(x)$ and 
the normalized density
$y(x) = n_{\rm H}(x)/n_0$ , $n_0$ being the mean hydrogen density,
are represented by their sampled values at equally spaced intervals
$\Delta x$, i.e. by the vectors \{$v_1, ... , v_k$\} and
\{$y_1, ... , y_k$\} with $k$ large enough to describe
the narrowest components of complex spectral lines.
For the ionization parameter as a function of $x$, we have
$U(x) = \hat{U}_0/y(x)$, with
$\hat{U}_0$ being the reduced mean ionization parameter
defined below.

\medskip\noindent
{\it Absorption system at $z_{\rm a} = 3.514$}.
A measurement of the primordial D/H in the $z_{\rm a} = 3.514$ system
has been recently made by Molaro et al. (1999).
They suggested that the blue wing of H\,{\sc i}~Ly$\alpha$ is contaminated
by D\,{\sc i} and evaluated a very low deuterium
abundance of D/H$ \simeq 1.5\times10^{-5}$ 
in the cloud with 
$N_{{\rm H}\,{\sc i}} = (1.23^{+0.09}_{-0.08})\times10^{18}$ cm$^{-2}$.
They considered, however, the derived D abundance as a lower limit
because their analysis was based on a simplified one-component 
VPF model
which failed to fit the red wing of the Ly$\alpha$ line as well as the
profiles of Si\,{\sc iii}, Si\,{\sc iv}, and  C\,{\sc iv} lines exhibiting
complex structures over approximately 100~km~s$^{-1}$ velocity range.
They further assumed that additional components would decrease the
H\,{\sc i} column density for the major component and, thus, would
yield a higher deuterium abundance.
Given the MCI method, we can test this assumption
since the MCI accounts self-consistently
for the velocity and density fluctuations.

Our aim is to fit the model spectra simultaneously to the observed
H\,{\sc i}, C\,{\sc ii}, C\,{\sc iv},
Si\,{\sc iii}, and Si\,{\sc iv} 
profiles.
In this case the mesoturbulent model requires 
the definition of a simulation
box for the six parameters~:
the carbon and silicon abundances,
$Z_{\rm C}$ and $Z_{\rm Si}$, respectively,
the rms velocity $\sigma_{\rm v}$ and 
density dispersion $\sigma_{\rm y}$,
the reduced total hydrogen column density
$\hat{N}_{\rm H} = N_{\rm H}/(1+\sigma^2_{\rm y})^{1/2}$, and
the reduced mean ionization parameter
$\hat{U}_0 = U_0/(1+\sigma^2_{\rm y})^{1/2}$.
For the model parameters the following boundaries were adopted~:
$Z_{\rm C}$ ranges from $10^{-6}$ to $4\times10^{-4}$,
$Z_{\rm Si}$ from $10^{-6}$ to $3\times10^{-5}$,
$\sigma_{\rm v}$ from 25 to 80 km~s$^{-1}$,
$\sigma_{\rm y}$ from 0.5 to 2.2,
$\hat{N}_{\rm H}$ from $5\times10^{17}$ to $8\times10^{19}$ cm$^{-2}$, and 
$\hat{U}_0$ ranges from $5\times10^{-4}$ to $5\times10^{-2}$.
We fix $z_{\rm a} = 3.51374$ (the value 
adopted by Molaro et al.) as a more or less arbitrary reference velocity
at which $v_j = 0$.

\begin{figure}
\vspace{-0.3cm}
\plotone{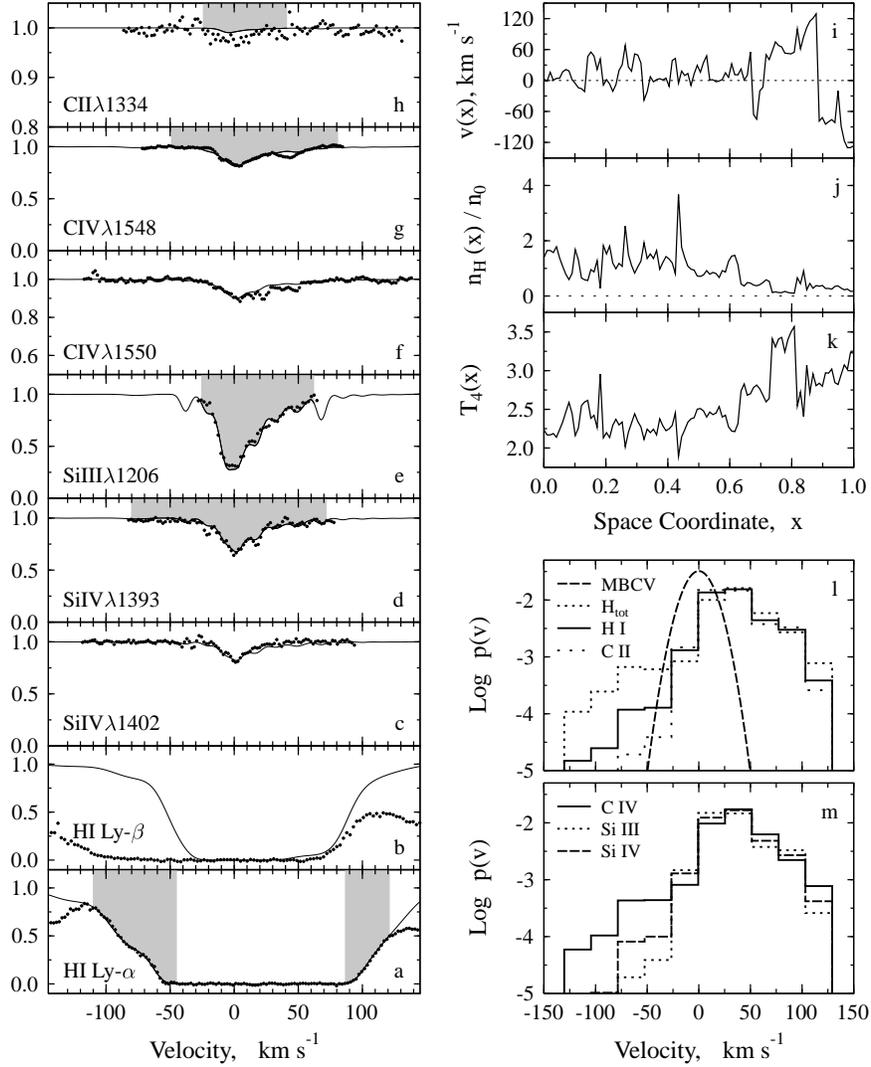}
\vspace{-4.0cm}
\caption{({\bf a}--{\bf h}). Observed normalized intensities (dots) 
and theoretical (solid curves) profiles of the absorption 
lines from the $z_{\rm a} = 3.514$ system (note different 
scales in panels {\bf h} and {\bf f}).
Shaded areas mark pixels which are critical to the MCI fit.
({\bf i}--{\bf k}). The corresponding MCI reconstraction of the
radial velocity, density, and kinetic temperature $T_4$ 
which is given in units
of $10^4$~K. ({\bf l},{\bf m}). The corresponding density-weighted
radial velocity distribution functions, $p(v)$, for the total
H$_{\rm tot}$ and neutral H\,{\sc i} hydrogen, C\,{\sc ii},
C\,{\sc iv}, Si\,{\sc iii}, and Si\,{\sc iv} as restored by the MCI.
For comparison, in panel {\bf l} the short dashed curve shows $p(v)$
adopted by Molaro et al. 
}
\end{figure}

Having specified the parameter space, we minimize the $\chi^2$ value.
The objective function includes 
those pixels which are critical to the fit. In Fig.~1, these pixels are 
marked by shaded areas.
In Fig.~1 (panels {\bf f} and {\bf c}),
the observed profiles of C\,{\sc iv}$\lambda1550$ and,
respectively, Si\,{\sc iv}$\lambda1402$
are shown together with the model spectra
computed with the
parameters derived from Ly$\alpha$, C\,{\sc ii}$\lambda1334$,
C\,{\sc iv}$\lambda1548$, Si\,{\sc iii}$\lambda1206$, and
Si\,{\sc iv}$\lambda1393$ fitting to illustrate the consistency.
For the same reason the Ly$\beta$
model spectrum is shown in panel {\bf b}
at the expected position.
All model spectra in Fig.~1 are drawn by continuous curves,
whereas filled circles represent observations (normalized fluxes).
The corresponding distributions of $v(x)$, $y(x)$, and $T(x)$
are shown in panels {\bf i}, {\bf j}, and {\bf k}.
The restored 
velocity field reveals a complex structure
which is manifested in non-Gaussian 
density-weighted velocity distribution
as shown in panels {\bf l}
and {\bf m} for the total hydrogen as well as for the
individual ions. 
We found that the radial velocity
distribution of H\,{\sc i} in the vicinity of 
$\Delta v \simeq - 100$ km~s$^{-1}$
may mimic the deuterium absorption and, thus, the asymmetric
blue wing of the hydrogen Ly$\alpha$ absorption may be
readily explained by H\,{\sc i} alone.
 
The median estimation of the model parameters gives
$N_{\rm H} = 5.9\times10^{18}$ cm$^{-2}$,
$N_{{\rm H}\,{\sc i}} = 5.3\times10^{15}$ cm$^{-2}$,
$U_0 = 1.6\times10^{-2}$,
$\sigma_{\rm v} = 51$ km~s$^{-1}$, and
$\sigma_{\rm y} = 1.1$.
The results were obtained with $k = 100$
and the correlation coefficients
$f_{\rm v} = f_{\rm y} = 0.95$
(for more details, see Levshakov, Agafonova, \& Kegel 2000a).

\begin{figure}
\plotfiddle{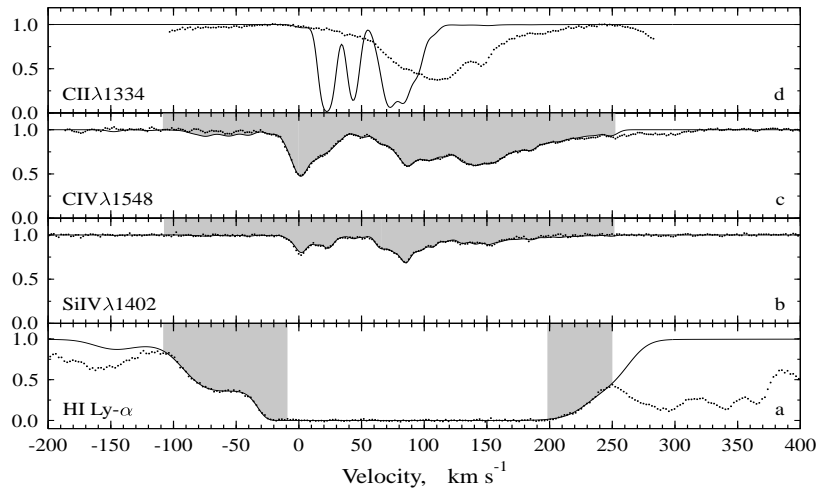}{8.0cm}{0}{80}{70}{-180}{45}
\vspace{-1.7cm}
\caption{({\bf a}--{\bf d}).
Observed normalized intensities (dots) 
and theoretical (solid curves) profiles of the absorption 
lines from the $z_{\rm a} = 3.378$ system
(shaded areas mark pixels used by the MCI).
The `H+D'-like absorption is explained by H\,{\sc i} alone.
An excellent common fit to H\,{\sc i}, Si\,{\sc iv}, and 
C\,{\sc iv} should, however, be rejected since the upper
panel demonstrates the inconsistency between the synthetic
spectrum of C\,{\sc ii} computed with the parameters derived
from the foregoing lines and the observed intensities. 
}
\end{figure}

The MCI allowed us to fit precisely not only the blue wing of the
saturated Ly$\alpha$ line but the red one as well.
We found that the actual neutral hydrogen column density
may be a factor of 250 lower than the value obtained by
Molaro et al. if one accounts for the velocity field structure.
Besides we did not confirm the extremely low metallicity
of [C/H] $\simeq - 4.0$, and [Si/H] $\simeq - 3.5$ reported by
Molaro et al. Our analysis yields
[C/H] $\simeq - 1.8$, and [Si/H] $\simeq - 0.7$. A similar silicon
overabundance has also been observed in halo (population~II) stars
(Henry \& Worthey 1999).

\medskip\noindent
{\it Absorption system at $z_{\rm a} = 3.378$}.
The following example illustrates how the realibility of the
inversion procedure can be controlled. We have chosen the 
$z_{\rm a} = 3.378$ system since at the position of the narrowest
C\,{\sc iv} subcomponent with $z_{{\rm C}\,{\sc iv}} = 3.37757$
and $b_{{\rm C}\,{\sc iv}} = 6.5$~km~s$^{-1}$ (see Ellison et al.)
one can see an `H+D'-like absorption in the blue wing of the
saturated Ly$\alpha$ line (Fig.~2a). 
The C\,{\sc iv}, and Si\,{\sc iv} profiles from this system were
treated by Ellison et al. separately. They found
$N^{\rm tot}_{{\rm C}\,{\sc iv}}  = (9.12^{+0.65}_{-0.61})\times10^{13}$ 
cm$^{-2}$ and  
$N^{\rm tot}_{{\rm Si}\,{\sc iv}} = (1.70 \pm 0.08)\times10^{13}$ cm$^{-2}$.   
For the neutral hydrogen, they estimated 
$N_{{\rm H}\,{\sc i}} = 2.8\times10^{15}$ cm$^{-2}$ and 
$b_{{\rm H}\,{\sc i}} = 78.6$ km~s$^{-1}$ (errors for both quantities
are greater than 30\%).

In this example, we assumed no deuterium absorption and tried to
force a common fit to the lines shown in panels {\bf a}--{\bf c}
(Fig.~2). Pixels used in the fitting are labeled by shaded areas. 
The MCI fit, shown by solid curves, looks excellent, and gives
$N_{{\rm H}\,{\sc i}} = 2.1\times10^{17}$ cm$^{-2}$,
$N_{{\rm C}\,{\sc iv}} = 1.0\times10^{14}$ cm$^{-2}$,
$N_{{\rm Si}\,{\sc iv}} = 2.6\times10^{13}$ cm$^{-2}$,
$\sigma_{\rm v} = 41.6$ km~s$^{-1}$, and 
$\sigma_{\rm y} = 1.3$.
The results were obtained with $k = 150$
and the correlation coefficients
$f_{\rm v} = f_{\rm y} = 0.97$.

The obtained MCI solution should, however, be rejected because
the synthetic spectrum of C\,{\sc ii} (solid curve in Fig.~2d),
computed with the parameters derived from the Ly$\alpha$, C\,{\sc iv},
and Si\,{\sc iv} fitting, differs significantly from the observed 
intensities (dots in Fig.~2d).
This example shows that 
we can always control the MCI results using additional
portions of the analysed spectrum.

Another issue of this example is that we can, in principle,
fit an `H+D'-like absorption by H\,{\sc i} alone even for
the systems with 
$N_{{\rm H}\,{\sc i}} \simeq 10^{17}$ cm$^{-2}$ and accompanying
metal lines. Examples of false deuterium identifications in systems
with 10 times lower neutral hydrogen column densities and
without supporting metal lines have been discussed in Tytler
\& Burles. Both cases stress the importance of the comprehensive
approach to the analysis of
each individual QSO system showing possible D absorption.

We may conclude that 
up-to-now deuterium was detected in only four QSO spectra
(Q~1937-1009, Q~1009+2956, Q~0130-4021, and Q~1718+4807) 
where $N_{{\rm H}\,{\sc i}}$ was measured with 
a sufficiently high accuracy. These measurements
are in concordance with ${\rm D/H} = (3-4)\times10^{-5}$.

{\it Acknowledgements}. 
This paper includes results obtained in collaboration with
I. I. Agafonova and W. H. Kegel.
The author is grateful to Ellison et al. 
for making their data available. 
I would also like to thank the conference
organizers for financial assistance.

\end{document}